\documentclass[prd,showpacs,preprintnumbers]{revtex4}
\usepackage[applemac]{inputenc}
\usepackage[T1]{fontenc}

\usepackage[dvips]{graphicx}

\usepackage{epsfig}
\usepackage{bbm}

\begin{document}
\preprint{UdeM-GPP-TH-07-163}
\preprint{arXiv:0710.5402[hep-th]}

\title{Spontaneous breaking of conformal invariance, solitons and gravitational waves in theories of conformally invariant gravitation}
\author{Jihène Bouchami$^{1}$}
\email{jihene.bouchami@umontreal.ca}  
\author{M. B.  Paranjape$^{1,2}$} 
\email{paranj@lps.umontreal.ca}
\affiliation{$^1$Groupe de physique des particules, D\'epartement de physique,
Universit\'e de Montr\'eal,
C.P. 6128, succ. Centre-ville, Montr\'eal, 
Qu\'ebec, Canada H3C 3J7 }
\affiliation{$^2$Center for Quantum Spacetime, Department of
Physics, Sogang University, Shinsu-dong \#1, Mapo-gu,
Seoul, 121-742, Korea}

\begin{abstract}
We study conformal gravity as an alternative theory of gravitation.  For conformal gravity to be phenomenologically viable requires that the conformal symmetry is not manifest at  the energy scales of the other known physical forces.  Hence we are required to find a mechanism for the spontaneous breaking of conformal invariance.  In this paper we study the possibility that conformal invariance is spontaneously broken due to interactions with conformally coupled matter fields.  The vacuum of the theory admits conformally non-invariant solutions corresponding to maximally symmetric space-times and variants thereof.  These are either de Sitter space-time or anti-de Sitter space-time in the full four space-time dimensions and we find new solutions corresponding to maximal symmetry restricted  to a lower dimensional sub-space.   We also consider  normalizable, linearized gravitational perturbations  around the anti-de Sitter
background. We show  to second order, that these gravitational fluctuations carry zero energy-momentum.    Finally we also show the possibility of domain wall solitons interpolating between the ground states of  spontaneously broken conformal symmetry that we have found.  These solitons necessarily require the vanishing of the scalar field. This offers a way of eschewing the recent suggestion and its consequences \cite{f} that the conformal symmetry could be quarantined to a sterile sector of the theory by choosing an appropriate field redefinition. 
\end{abstract}


\pagenumbering{arabic}

\maketitle

\section{Introduction}
Einsteinian general relativity is the widely accepted
classical theory of gravity.  Enjoying great success at the solar
system scale, it has also been able to explain many  stellar, galactic and
cosmological phenomena \cite{weinberg}.  However, the problem of flat rotational
curves of spiral galaxies and the supplementary deflexion of light
by galaxies and clusters rule out the validity of the standard gravitational
law unless dark matter exists \cite{sanders}.  The introduction of dark energy has been necessary to explain
the observed acceleration of the universe \cite{turner}. In addition, the Einstein theory 
does not seem to be  consistent as quantum
theory of gravity \cite{deser}.
For all these reasons,  Einsteinian general relativity quite possibly may not be the correct theory of gravity. Indeed at the present epoch, the study of alternative theories of gravity has reached an apogee of different possibilities, {\it viz}, Born-Infeld gravity, Brans-Dicke gravity, Lovelock gravity, MOND, $f(R)$ gravity  and conformal gravity \cite{l,bd,lo,Milgrom,bekenstein,mann1}.

Conformal gravity is a generally covariant theory of gravitational interactions, which benefits from an additional infinite dimensional invariance, invariance under local conformal transformations.   Perturbative calculations indicate that it is asymptotically free and power counting renormalizable \cite{stelle}. It is also could define a unitary, ghost free quantum field theory due to non-perturbative reorganization of the asymptotic physical states \cite{ft,t}.  Conformal gravity has been suggested as an alternative to the
standard theory. It was originally realized by Weyl in the
early age of general relativity. To introduce a gauge symmetry for
gravity, Weyl proposed that the gravitational action has to be made up
of fourth order derivative terms to make it invariant under local
conformal transformations of the metric \cite{schmidt}.  

But conformal gravity overcomes other challenges as well.  Mannheim and Kazanas showed
that the Weyl theory is capable of explaining the observed flat rotation curves at galactic and extragalactic scales without the need
of dark matter \cite{mann1}.  Moreover, it was proposed a
conformal cosmological model solving the flatness and horizon
problems without predicting an inflationary universe \cite{mann5}.  A difficulty does arise when the deflection of light is studied. It was noticed that the deflection of light is actually diminished, in conflict with observations, by the linear potential that was found and used in the analysis by Mannheim and Kazanas \cite{mann1} and which is so crucial to explain the the flat galactic rotation curves \cite{ep}.  A possible solution was offered by \cite{emp}, where it was observed that although the solution for the flat galactic rotation curves is crucially dependent on the the choice of conformal gauge, the deflection of light, a massless particle, is not.  Conformal transformations leave the null geodesics invariant.  Hence one can choose the linear potential so that the deflection of light is augmented and then choose the conformal gauge so that the galactic rotation curves are explained.  For details see \cite{emp}. 

However, this leads us to the most serious defect of the conformally invariant theory. The physical existence of massive particles implies non invariance under conformal transformations.  There is an evident lack of conformal invariance in the observed physical phenomena, the tangible world has a definite scale.  Thus for conformal gravity to be a realistic alternative theory of gravitation there must be a mechanism by which the conformal invariance is either broken, at least at the scales of physics that we have been able to heretofore probe, or somehow transferred to a hidden sector, so that it is not manifest.  The latter possibility has been implemented in Ref. \cite{f} where it is explained that it is indeed possible for a theory to have an invariance under conformal transformations without contradicting the fact that conformal invariance is not seen experimentally.

In this paper, however, we study the alternative of spontaneous breaking of conformal invariance in the conformally invariant theory of gravitation coupled to scalar matter.  We focus on anti-de Sitter space-time, which is a vacuum solution of conformal gravity that breaks conformal invariance and  can offer an explanation for repulsive gravitation.\cite{mannheim4}.  We study the spectrum of its small gravitational  fluctuations.  Using the conformal invariance, we map the problem to the simpler one corresponding to the study of fluctuations about flat Minkowski space-time.  We find that the fluctuations which are fourier decomposable, carry zero energy and momentum.  It has been shown for the full non-linear theory, \cite{bhs}, that asymptotically flat space-times in conformally invariant gravity have exactly zero energy and momentum.  Our result fits well with this theorem.  

It has also been recently pointed out that in the conformal gravity theory with a conformally coupled scalar field , it is possible to make a change of variables to relegate the conformal symmetry to a sterile, disjoint sector of the theory, that does not interact with the rest of the fields which do not exhibit any conformal symmetry \cite{f}.  However, a crucial assumption required that the scalar field never vanishes.  Here we study domain wall type solitons in our conformally invariant theories with spontaneous breaking of conformal invariance.  We find the possibility of domain wall solitons which interpolate between different vacua of the spontaneously broken theory.  These solitons necessarily pass through zeros of the scalar field, essentially defining the location of the domain wall.  These solitons are complimentary to the magnetic monopole type solitons that we previously found \cite{efp}, which also require the vanishing of the scalar field at the location of the monopole.  The existence of such configurations and their import to the full quantum theory then denies the conclusions of the analysis made in \cite{f}.

However it is possible that the initial value problem for the theory is well posed only on a subset of field configurations.  There is some reason to suspect that the initial value problem is ill posed exactly on the set of fields where the scalar field vanishes \cite{fl}.  Consequently, the theory should be considered as provisional until this question has been resolved.  

\section{Conformal gravity}

It has been suggested that gravity can be described by a fourth
order derivative theory based on both covariance principle and
conformal symmetry \cite{mann1}. This theory is called conformal
gravity whose action has the form
\begin{eqnarray}
I_{W}&=&-\alpha\int d^{4}x
\sqrt{-g}~C_{\mu\nu\sigma\tau}C^{\mu\nu\sigma\tau}\label{weylaction1}\\
&=&-\alpha\int
d^{4}x \sqrt{-g}~\Big(R_{\mu\nu\sigma\tau}R^{\mu\nu\sigma\tau}-2R_{\mu\nu}R^{\mu\nu}+\frac{1}{3}R^{2}\Big )\label{weylaction2}\\
&=&-2\alpha\int
d^{4}x \sqrt{-g}\Big (R_{\mu\nu}R^{\mu\nu}-\frac{1}{3}R^{2}\Big)\label{weylaction3},
\end{eqnarray}
where $C_{\mu\nu\sigma\tau}$ is the Weyl tensor, given by
\begin{equation}
C_{\mu\nu\sigma\tau}=R_{\mu\nu\sigma\tau}-\frac{1}{2}\left(g_{\mu\sigma}R_{\nu\tau}-g_{\mu\tau}R_{\nu\sigma}-g_{\nu\sigma}R_{\mu\tau}+g_{\nu\tau}R_{\mu\sigma}\right) +\frac{R}{6}\left(g_{\mu\sigma}g_{\nu\tau}-g_{\mu\tau}g_{\nu\sigma}\right) ,
\end{equation}
$R_{\mu\nu\sigma\tau}$ is the Riemann tensor,  $R_{\mu\nu}$ is the  Ricci tensor, $R$ is the curvature scalar and $\alpha$ is a 
dimensionless parameter.  Equation (\ref{weylaction2}) follows directly from the definition of the Weyl tensor, while equation (\ref{weylaction3}) follows from equation (\ref{weylaction2}) and the expression for the Euler characteristic, 
$$
\chi = \sqrt{-g}~\Big(R_{\mu\nu\sigma\tau}R^{\mu\nu\sigma\tau}-4R_{\mu\nu}R^{\mu\nu}+R^{2}\Big )
$$ 
a total derivative, which is then used to eliminate the terms involving the Riemann tensor \cite{mannheim4}.  

The sign of $\alpha$ can be fixed by insisting that the Euclidean action for small fluctuations be positive.  This condition is absolutely necessary if the Feynman functional integral is to have any chance of defining the corresponding quantum theory.  The Minkowski Feynman functional integral is given by 
$$
{\cal Z}=\int{\cal D\phi}e^{{i\cal I}_{Mink.}}
$$
an integral over the space of field configurations weighted by the exponential of $i$ times the Minkowski action is not very well defined.  The Euclidean path integral
$$
{\cal Z}=\int{\cal D\phi}e^{{-\cal I}_{Eucl.}}
$$
is much better defined, however it is crucial that ${\cal I}_{Eucl.}$ be non-negative.  The Minkowski  functional integral and the corresponding quantum amplitudes are actually obtained by anaytically continuing back from Euclidean space.
In usual field theory, that is second order in time derivatives, the Minkowski action is of the form
$$
{\cal I}_{Mink.}=\sigma\int d^4 x~( T-V)\sim \sigma\int d td^3 x~(\frac{1}{2} (\dot\phi)^2-V).
$$
The continuation to Euclidean space requires the replacement $t\rightarrow -i\tau$, which gives
$$
i{\cal I}_{Mink.}\rightarrow -{\cal I}_{Eucl.}=i(-i)(-)\sigma\int d \tau d^3 x~(\frac{1}{2} (\frac{\partial\phi}{d\tau})^2+V).
$$
Evidently the term of highest temporal derivatives (two) changes sign under the analytic continuation, and hence combines with the potential term, giving an overall minus sign and a non-negative Euclidean action.  Thus the coefficient $\sigma$ must be positive and normally we take $\sigma =1$.  

For a theory with higher temporal derivatives (four), such as the theory of conformal gravity that we are studying here, the Minkowski action has the form
\begin{equation}
{\cal I}_{Mink.}=-\alpha\int d^4 x~( T+V)\sim -\alpha \int d td^3 x~(\frac{1}{2} (\ddot\phi)^2+V)\label{action}
\end{equation}
where we have switched the sign of the potential taking out an overall minus sign, as we are anticipating that $\alpha$ is positive with the choice of the kinetic term as written.
This form is certainly true for the dynamics of the perturbative fluctuations and is amenable to continuation to Euclidean space, however, we must note that the analytic continuation of a general curved Lorentzian manifold to Euclidean space is not necessarily straightforward.  Now continuation of (\ref{action}) to Euclidean space yields
$$
i{\cal I}_{Mink.}\rightarrow -{\cal I}_{Eucl.}=i(-i)(-)\alpha\int d \tau d^3 x~(\frac{1}{2} (\frac{\partial^2\phi}{d\tau^2})^2+V),
$$
thus requiring $\alpha$ to be positive.  

The action (\ref{weylaction1}) is invariant under local
conformal transformation of the metric
\begin{equation}\label{trans}
g_{\mu\nu}(x)\rightarrow\Omega^{2}(x)g_{\mu\nu}(x),
\end{equation}
where $\Omega(x)$ is a real, continuous and non-vanishing
function.  The variation of equation (\ref{weylaction2}) or equivalently equation (\ref{weylaction3}) gives the matter free
gravitational field equations $W_{\mu\nu}=0$, where
\begin{equation}
g_{\mu\gamma}g_{\nu\beta}\frac{\delta I}{\delta g_{\gamma\beta}}=-2\alpha \sqrt{-g}W_{\mu\nu}
\end{equation}
and
\begin{eqnarray}\label{tenswmunu}
W_{\mu\nu}=-\frac{1}{6}g_{\mu\nu}R^{;\sigma}_{~~;\sigma}
+\frac{2}{3}R_{;\mu;\nu}+R_{\mu\nu~;\sigma}^{~~;\sigma}
-R_{\mu~;\nu;\sigma}^{~\sigma}-R_{\nu~;\mu;\sigma}^{~\sigma}\nonumber\\
-2R_{\mu\sigma}R_{\nu}^{~\sigma}+\frac{1}{2}g_{\mu\nu}R_{\sigma\tau}R^{\sigma\tau}
+\frac{2}{3}RR_{\mu\nu}-\frac{1}{6}g_{\mu\nu}R^2.
\end{eqnarray}
$W_{\mu\nu}$, called the Bach tensor, is automatically  covariantly conserved (for any choice of the metric), which corresponds to the Bianchi identities, 
$$
W_{\mu ~~;\nu}^{~\nu}=0
$$
due to the invariance of the action under general coordinate transformations.  It is  also automatically trace free  (for any choice of the metric)
$$
W^{\mu}_{~~\mu}=0
$$
due to the invariance of the action under local conformal transformations.  The solutions of the equations of motion $W_{\mu\nu}=0$ correspond to the vacuum configurations of conformal gravity \cite{mann1}.

Introducing a scalar field $\varphi(x)$ in this theory and maintaining the conformal invariance, requires the well-known action
\begin{equation}\label{actionfield}
I_{M}=\int d^{4}x \sqrt{-g}\Big
[\frac{1}{2}\partial_{\mu}\varphi\partial^{\mu}\varphi+\frac{1}{12}R\varphi^2-\lambda
\varphi^4 \Big ],
\end{equation}
where $R$ is the scalar curvature and $\lambda$ is a dimensionless
coupling constant. Notice that there is no mass term.  Under the conformal
transformation of the metric (\ref{trans}) and  the corresponding transformation of the scalar field
\begin{equation}\label{transfield}
\varphi(x)\rightarrow\Omega^{-1}(x)\varphi(x)
\end{equation}
the scalar action is invariant.  The lack of conformal invariance of the kinetic term is exactly cancelled by that of the $R\varphi^2$ term.  

The variation of the action (\ref{actionfield}) with respect to the metric defines the
energy-momentum tensor of the scalar field
\begin{eqnarray}\label{tenseurtmunu}
T_{\mu\nu}&=&\frac{2}{3}\partial_{\mu}\varphi\partial_{\nu}\varphi
-\frac{1}{6}g_{\mu\nu}\partial_{\sigma}\varphi\partial^{\sigma}\varphi\nonumber
-\frac{1}{3}\varphi
D_{\nu}\partial_{\mu}\varphi\\
&+&\frac{1}{3}g_{\mu\nu}\varphi D_{\sigma}\partial^{\sigma}\varphi
+\frac{1}{6}\varphi^2(R_{\mu\nu}-\frac{1}{2}g_{\mu\nu}R)+g_{\mu\nu}\lambda
\varphi^4,
\end{eqnarray}
Noting the normalization of $W_{\mu\nu}$ in equation (\ref{tenswmunu})
the full gravitational motion equations are simply
\begin{equation}\label{eqs}
2 \alpha W_{\mu\nu}= \frac{1}{2}T_{\mu\nu}.\label{12}
\end{equation}
The two  tensors in Equation(\ref{12}) are symmetric, traceless and independently, covariantly conserved. Under conformal transformations (\ref{trans}) and (\ref{transfield}), they
transform as $W_{\mu\nu}\rightarrow\Omega^{-4}W_{\mu\nu}$ and
$T_{\mu\nu}\rightarrow\Omega^{-4}T_{\mu\nu}$.  The scalar field equation of motion is obtained by varying with respect to $\varphi$, yielding
\begin{equation}
\frac{1}{\sqrt{-g}}\partial_\mu\sqrt{-g}\partial^\mu\varphi -\frac{1}{6}R\varphi+4\lambda\varphi^3\label{phieq}=0
\end{equation}

\section{Spontaneous breaking of conformal symmetry}
Vacuum solutions of the field equations satisfy
$$
W_{\mu\nu}=0.
$$
Evidently, Minkowski space-time, $g_{\mu\nu}=\eta_{\mu\nu}$, is a solution. Then, using the conformal invariance, every metric that is conformally related to Minkowski space-time,  $g_{\mu\nu}=\Omega^2(x) \eta_{\mu\nu}$, is also a solution, where $\Omega(x)$ is an arbitrary but sufficiently smooth, non-vanishing function of the coordinates.  Taking into account  the scalar field, a full vacuum solution corresponds to $g_{\mu\nu}=\eta_{\mu\nu}, \varphi=0$.  Every conformal transform of this solution is also a solution. 

These solutions do not break conformal invariance, no mass scale is generated.  However there do exist solutions which spontaneously break the conformally invariance.  We do not establish an exhaustive classification of such solutions here, but such an exercise would be very interesting.  

\subsection{Maximally symmetric vacuum solutions with broken conformal invariance\label{mss}}

Maximally symmetric space-times have been classified for four dimensional Lorentzian\hyphenation{Lo-rentz-i-an} metric space-times \cite{weinberg}.   We can use these geometries to establish a class of vacuum solutions that spontaneously break the conformal invariance.  These solutions have been first found by Mannheim\cite{mannheim4}.  Maximally symmetric four dimensional space-times are simply given by either de Sitter space-time, anti-de Sitter space time or Minkowski space-time.  They are characterized by an arbitrary, constant, curvature scalar $R$, a Ricci tensor given by $R_{\mu\nu}=\frac{R}{4}g_{\mu\nu}$ and a Riemann tensor given by $R_{\mu\nu\sigma\tau}=\frac{R}{12}(g_{\mu\sigma}g_{\nu\tau}-g_{\mu\tau}g_{\nu\sigma})$.  The Bach tensor vanishes for any value of the (constant) curvature scalar, since all of these geometries are conformally related to Minkowski space.  Hence the matter free gravitational field equations $W_{\mu\nu}=0$ are satisfied, and this is why we call these solutions vacuum solutions.  The scalar field equation (\ref{phieq}) and the gravitational field equations (\ref{eqs}),  taking into account  $W_{\mu\nu}=0$ and the assumption that $\varphi$ is constant, give
\begin{eqnarray}
-\frac{1}{6}R\varphi+4\lambda\varphi^3&=&0\label{f1}\\
T_{\mu\nu}=
\frac{1}{6}\varphi^2(R_{\mu\nu}-\frac{1}{2}g_{\mu\nu}R)+g_{\mu\nu}\lambda
\varphi^4&=&0.\label{f2}
\end{eqnarray}
Upon replacement $R_{\mu\nu}=\frac{R}{4}g_{\mu\nu}$, equation (\ref{f2}) somewhat serendipitously becomes identical to equation (\ref{f1}) except for being multiplied by the factor $g_{\mu\nu}\varphi$, allowing for the simultaneous solutions of both equations by, $\varphi=0$ or $\varphi=\pm\sqrt{R\over 24\lambda}$.  The solution $\varphi=0$ has already been discussed and does not break conformal invariance however the second solution exists for positive constant curvature space-times and spontaneously breaks the conformal invariance by generating a mass scale.   Such space-times correspond to anti-de Sitter space-time in our convention.  The solutions for different values of the scalar curvature, which serves as a modulus parameter,  are related to each other by conformal transformations.  Analysis of the small fluctuations about these solutions will be done in the later sections.  

The calculation of the possibility of quantum tunneling between the different vacua is well beyond the scope of this paper, however it is an extremely interesting subject.  The non conformally invariant vacua, $\varphi\ne0$,  actually correspond to a negatively infinite total action since the minimum of the effective scalar field potential, at constant values of the curvature, occurs at a non-zero value of the potential.  On the other hand, the simple $\varphi=0$ vacuum corresponds to zero total action which lend credence to the conclusion that the symmetry breaking solutions are the stable ground states.  The gravitational part of the action for a maximally symmetric space-time does vanish, indeed the expressions given by equations (\ref{weylaction1}) and (\ref{weylaction2}) do vanish, thus these cannot compensate for the infinite action in the scalar field part.  We note, however that the action given in equation (\ref{weylaction3}) does not vanish, and is in fact infinite.  The explanation for this is given by the fact that although the Euler characteristic is a total divergence, its value for a maximally symmetric space-time is infinite, hence the value of the action given by (\ref{weylaction3}) must also be infinite to compensate.  This fact can render analyses, such as those used in Derrick's theorem, where the stability or instability of field configurations is determined by explicitly looking at the behaviour of the action under changes of scale, problematic.  We repeat here in order to underline, that the field equations obtained by the variation of any of the expressions for the action, are of course identical, because the variation that is considered is local.  

\subsection{Non-maximally symmetric solutions with broken conformal invariance\label{nmss}}

There exist other, essentially vacuum solutions to the field equations (\ref{eqs}) and (\ref{phieq}) which do not require that the Bach tensor vanishes.  The solutions are essentially vacuum solutions since the curvature scalar is a constant as is the scalar field, but the space-time is not maximally symmetric.  Consider the ansatz
\begin{equation}
d\tau^2=(1+B(x))dt^2-\frac{1}{(1+B(x))}dx^2-dy^2-dz^2.\label{ansatz}
\end{equation}
The coordinates $(x,y,z,t)$ each vary from $-\infty$ to $\infty$.  The metric describes a space-time with the topology ${\mathbf{M}^2} \times{\mathbf R}^2$ a two dimensional Lorentz manifold cross the flat two dimensional Euclidean space.  The curvature scalar is given by $R(x)=B^{\prime\prime}(x)$.  $B(x)=\frac{R}{2} x^2$ yields a constant curvature space-time for $R$ a constant, the curvature scalar is evidently given by $R$.  The two dimensional Lorentz manifold is exactly two dimensional de Sitter or anti-de Sitter space-time.  This ansatz for the metric together with the ansatz that the scalar field is a constant yields a solution of the field equations.  This solution is fundamentally different from the solution of the section (\ref{mss}).  Here the Bach tensor does not vanish, and hence a solution exists only for a critically coupled system in terms of the parameters $\alpha$ and $\lambda$ and for a specific value of the scalar field.  The solutions for different values of the scalar curvature are not related to each other by conformal transformations, nevertheless, the value of the scalar curvature still serves as a modulus parameter. 

Nominally, eleven field equations (one symmetric two tensor and the scalar field equation) for two functions $B(x)$ and $\varphi(x)$ seems hopeless, however the symmetries and the constraints satisfied by the Bach tensor and the energy-momentum tensor actually reduces to exactly two equations.  Such an analysis was first carried out in \cite{mannheim8} when solutions of the Kerr-Reissner-Nordstrom were found.  With the ansatz (\ref{ansatz}) and $\varphi$ constant, the gravitational field equations reduce to the four diagonal equations together with the scalar field equation
\begin{eqnarray}
2\alpha W_{00}&=&\frac{1}{2}T_{00}\label{1}\\
2\alpha W_{11}&=&\frac{1}{2}T_{11}\label{2}\\
2\alpha W_{22}&=&\frac{1}{2}T_{22}\label{3}\\
2\alpha W_{33}&=&\frac{1}{2}T_{33}\label{4}\\
-\frac{1}{6}R\varphi+4\lambda\varphi^3&=&0.\label{5}
\end{eqnarray}
The translational symmetry in spatial directions $y$ and $z$ yields that $W_{22}=W_{33}$ and $T_{22}=T_{33}$.  Using this in the equations for the independent tracefree nature of the Bach tensor and the energy momentum tensor gives us the constraints 
$$ 
g^{00}W_{00}+g^{11}W_{11}-2W_{22}=0
$$ and 
$$ 
g^{00}T_{00}+g^{11}T_{11}-2T_{22}=0.
$$  
The first equation is directly valid, while the second reduces to the scalar field equation (\ref{5}).  Hence equations (\ref{3}) and (\ref{4}) are automatic if the remaining equations  (\ref{1}) (\ref{2}) and (\ref{5}) are satisfied.   The covariant conservation of both tensors yields 
$$
W_{\alpha~;0}^{~0}+W_{\alpha~;1}^{~1}=0
$$ 
and 
$$
T_{\alpha~;0}^{~0}+T_{\alpha~;1}^{~1}=0,
$$
for $\alpha=0,1$.  We have used the fact that the space-time is flat in the $(y,z)$ directions hence the explicit derivatives and the connection coefficients simply vanish for these directions.   Calculating the $(0,1)$ direction connections gives
\begin{equation}
\Gamma^0_{\alpha\beta}=\frac{g_{00}^\prime}{2g_{00}}\Big(\begin{array}{cc}0&1\\1&0\end{array}\Big) ,\quad \Gamma^1_{\alpha\beta}=\Big(\begin{array}{cc}g_{00}g_{00}^\prime/2&0\\ 0&{-g_{00}^\prime}/{2g_{00}}\end{array}\Big)
\end{equation}
where $\alpha, \beta$ take on the values 0 and 1.  It is then easy to see that $W_{00}$ depends linearly on $W_{11}$ and $T_{00}$ on $T_{11}$.  Thus the equation (\ref{1}) is automatic if the remaining two equations  (\ref{2})and  (\ref{5}) are satisfied.  Thus the system reduces to the two equations
\begin{eqnarray}
2\alpha W_{11}&=&\frac{1}{2}T_{11}\label{6}\\
-\frac{1}{6}R\varphi+4\lambda\varphi^3&=&0,\label{7}
\end{eqnarray}
which are to be solved for two independent constants $R$ and $\varphi$.  Explicitly the equations are given by
\begin{eqnarray}
\alpha (B'')^2/3 &=&\lambda\varphi^4\label{8}\\
-\frac{1}{6}(B'')\varphi+4\lambda\varphi^3&=&0.\label{9}
\end{eqnarray}
where we have multiplied through by a factor of $1+B$ in equation (\ref{8}).  A non-zero constant solution for $\varphi$ requires $B(x)=Rx^2/2$ (where $R$ is the (constant) curvature scalar) which yields the condition on the couplings
$$
\alpha  =1/192\lambda .
$$
On first sight this condition does not fix the value of the constant scalar curvature $R$, it could be positive or negative, corresponding to two dimensional anti-de Sitter or de Sitter space-time respectively.  Only flat Minkowski space-time is apriori excluded, since we have assumed $R\ne 0$.  However, de Sitter space-time is actually excluded by equation (\ref{9})  which yields
$$
\varphi^2=\frac{B^{\prime\prime}}{24\lambda}
$$
hence $R=B''$ must be positive and choosing anti-de Sitter space-time.  This is exactly the same value as the solution for the scalar field in subsection(\ref{mss}) since the scalar field equation and its solution, for constant scalar field and constant scalar curvature, is unique.   Our solution is new and involves in a non-trivial way the fourth order Weyl action, in contradistinction to the vacuum solutions corresponding to maximally symmetric anti-de Sitter space-time that we found in subsection (\ref{mss}).

\section{Fluctuations about the vacuum solutions}

In this section we study the behaviour of gravitational fluctuations about the anti-de Sitter vacuum solution that we have analyzed in Section (\ref{mss}).  We do not provide a detailed analysis of the general problem of perturbation theory, it has been studied extensively in the past, see \cite{ft} and references therein.  Indeed it is commonly felt that the linearized perturbation theory gives rise to run-away solutions, which correspond to ghost-like states and ruin the unitarity of the corresponding quantum field theory, see for example \cite{bw}.  This has to be examined in the context of the following articles, where it has been shown that a theory fourth order in derivatives can define a power counting renormalizable theory \cite{stelle}, is asymptotically free \cite{ft2}, is unitary in the large $N$ limit \cite{t3}, is unitary in a Hamiltonian quantization where no ghosts are seen to leading order in strong coupling \cite{k}, satisfies the zero total energy theorem \cite{bhs},  and in a Euclidean lattice formulation has a positive norm Hilbert space and positive Hamiltonian which satisfies reflection positivity and hence allows for construction of a unitary theory from the Osterwalder Schrader \cite{os} reconstruction theorem \cite{t2}.  In a very recent development, Bender and Mannheim \cite{bm} in a recent preprint have shown that a simple fourth order theory, the Pais-Uhlenbeck oscillator is perfectly free of ghosts if it is properly quantized with regards to its $PT$ symmetry.  All of these indicators suggest that the unitarity of the full theory is not impinged upon by the unitarity, or lack therof, of the linearized perturbation theory.  The actual asymptotic states, which are the relevant states with respect to the unitarity of the $S$-matrix can have little or nothing to do with the states of the linear theory.  For example, in QCD, the linear theory sees quarks and gluons, however, the asymptotic states are explicitly colour-free bound-state combinations of these, the baryons and the mesons.  In \cite{k} it is found that any ghost states of the linear perturbation theory are confined due to the constraints of the theory while the graviton emerges a string-like solution to the Dirac constraints of the conformal invariance of the theory, at strong coupling.  

In the analysis below, we do not offer any conclusive evidence as to the unitarity of the quantum theory of conformal gravity.  We study normalizable gravitational fluctuations, which do correspond to gravitational waves with respect to anti-de Sitter background vacuum solution of conformal gravity.  We find the interesting result that these fluctuations carry no energy or momentum, to second order perturbation theory.

\subsection{Gravitational waves in anti-de Sitter space-time}

We study the linearized gravitational perturbation around the
anti-de Sitter background in conformal gravity. The metric takes the form
\begin{equation}\label{metric perturbation}
\widetilde{g}_{\mu\nu}=\gamma_{\mu\nu}+\widetilde{h}_{\mu\nu}=\Omega^{2}(\eta_{\mu\nu}+h_{\mu\nu})=\Omega^{2}g_{\mu\nu},
\end{equation}
where $\gamma_{\mu\nu}$ and $\eta_{\mu\nu}$ are respectively anti-de
Sitter and Minkowski metrics, $\widetilde{h}_{\mu\nu}$ and
$h_{\mu\nu}$ are small perturbations respectively around
$\gamma_{\mu\nu}$ and $\eta_{\mu\nu}$ and $\Omega(x)$ is the
conformal factor relating the two metrics. We note that the indices
of $h_{\mu\nu}$, $\widetilde{h}_{\mu\nu}$ and all zero order tensors
are raised and lowered with $\gamma_{\mu\nu}$ metric, however for
the remainder of tensors we use the full metric $g_{\mu\nu}$ (or
$\widetilde{g}_{\mu\nu}$).

To study perturbation around anti-de Sitter metric, it is more
practical to transform the work on the flat space-time since
$g_{\mu\nu}$ and $\widetilde{g}_{\mu\nu}$ are conform one to
another. So for $T_{\mu\nu}=0$ (in absence of sources), equation
(\ref{eqs}) gives
\begin{equation}\label{sansT}
\widetilde{W}_{\mu\nu}=\Omega^{-4}W_{\mu\nu}=0,
\end{equation}
where $\widetilde{W}_{\mu\nu}$ and $W_{\mu\nu}$ are Bach tensors
respectively of anti-de Sitter and Minkowski metrics. To first order
in $\widetilde{h}_{\mu\nu}$ and $h_{\mu\nu}$, we have
\begin{equation}\label{worder1}
\widetilde{W}_{\mu\nu}^{(1)}=\Omega^{-4}W_{\mu\nu}^{(1)}=0,
\end{equation}
where
\begin{eqnarray}\label{wmink}
W_{\mu\nu}^{(1)}=-\frac{1}{6}\eta_{\mu\nu}\partial_{\rho}\partial^{\rho}(\eta^{\lambda\rho}R_{\lambda\rho}^{(1)})
+\frac{2}{3}\partial_{\mu}\partial_{\nu}(\eta^{\lambda\rho}R_{\lambda\rho}^{(1)})\nonumber\\
+\partial_{\rho}\partial^{\rho}
R_{\mu\nu}^{(1)}-\partial_{\beta}\partial_{\mu}(\eta^{\alpha\beta}R_{\nu\alpha}^{(1)})
-\partial_{\beta}\partial_{\nu}(\eta^{\alpha\beta}R_{\mu\alpha}^{(1)})=0.
\end{eqnarray}
Since
\begin{equation}\label{rmunuorder1}
R_{\mu\nu}^{(1)}=\frac{1}{2}\Big
(\partial_{\beta}\partial_{\mu}h_{\nu}^{\beta}
+\partial_{\beta}\partial_{\nu}h_{\mu}^{\beta}
-\partial_{\mu}\partial_{\nu}h_{\beta}^{\beta}
-\partial_{\rho}\partial^{\rho} h_{\mu\nu}\Big ),
\end{equation}
So, $h_{\mu\nu}$ is a solution of the equation
\begin{eqnarray}\label{eqofh}
-\frac{1}{2}(\partial_{\rho}\partial^{\rho})^{2}h_{\mu\nu}+\frac{1}{6}\eta_{\mu\nu}(\partial_{\rho}\partial^{\rho})^{2}h_{\beta}^{\beta}-\frac{1}{6}\eta_{\mu\nu}\partial_{\rho}\partial^{\rho}
(\eta^{\lambda\rho}\partial_{\alpha}\partial_{\lambda}h_{\rho}^{\alpha})-\frac{1}{6}\eta^{\nu\beta}\partial_{\mu}\partial_{\beta}\partial_{\rho}\partial^{\rho}
h_{\alpha}^{\alpha}\nonumber\\
-\frac{1}{3}\eta^{\nu\beta}\eta^{\lambda\rho}\partial_{\mu}\partial_{\beta}\partial_{\alpha}\partial_{\lambda}h_{\rho}^{\alpha}
+\frac{1}{2}\eta^{\nu\rho}\partial_{\rho}\partial_{\beta}\partial_{\rho}\partial^{\rho}
h_{\mu}^{\beta}+\frac{1}{2}\eta^{\beta\lambda}\partial_{\lambda}\partial_{\mu}\partial_{\rho}\partial^{\rho}
h_{\nu\beta}=0.
\end{eqnarray}
This purely fourth order derivative equation cannot be simplified
without fixing gauge conditions. To fix the invariance of the
theory under coordinate transformation we choose the harmonic gauge
$g^{\mu\nu}\Gamma^{\lambda}_{\mu\nu}=0$ (where
$\Gamma^{\lambda}_{\mu\nu}$ is the Christoffel coefficient relied to
$g_{\mu\nu}$ ) that can be written at first order
\begin{equation}\label{gharm}
\partial_{\mu}h_{\nu}^{\mu}=\frac{1}{2}\partial_{\nu}h_{\mu}^{\mu}.
\end{equation}
Furthermore, since conformal gravity has an additional conformal symmetry, $h_{\mu\nu}$ can be further restricted by a second
gauge choice. If we perform an infinitesimal conformal
transformation which the factor $\Omega(x)$ has the form
$(1+\epsilon(x))$ where $\epsilon(x)$ is a function having the same
order of magnitude as $h_{\mu\nu}$, then the metric $g_{\mu\nu}$
transforms as
\begin{equation}
g_{\mu\nu}\rightarrow
g_{\mu\nu}'=(1+\epsilon)(\eta_{\mu\nu}+h_{\mu\nu})=\eta_{\mu\nu}+h_{\mu\nu}+\epsilon
\eta_{\mu\nu}+O(h^2)\simeq \eta_{\mu\nu}+h'_{\mu\nu}\nonumber,
\end{equation}
where $h'_{\mu\nu}=h_{\mu\nu}+\epsilon \eta_{\mu\nu}$. If we take
the trace of this equation, we find
$h_{\mu}^{'\mu}=h_{\mu}^{\mu}+4\epsilon$. Then, we can choose
$\epsilon=-\frac{1}{4}h_{\mu}^{\mu}$ to yield $h_{\mu}^{'\mu}=0$.

If we choose to place in the space-time described by the metric
$g'_{\mu\nu} \simeq \eta_{\mu\nu}+h'_{\mu\nu}$, we use the gauge
conditions (we can omit the ')
\begin{equation}\label{cond1}
\partial_{\mu}h_{\nu}^{\mu}=0
\end{equation}
and
\begin{equation}\label{cond2}
h_{\mu}^{\mu}=0.
\end{equation}
With these conditions, the gravitational perturbation $h_{\mu \nu}$ obeys the equation
\begin{equation}\label{eqpert}
(\partial_{\rho}\partial^{\rho})^{2}h_{\mu\nu}=0.
\end{equation}
The general solution of this equation can be written as a linear
superposition of plane waves. For simplicity, we consider one mode
\begin{equation}\label{mode}
h_{\mu\nu}=e_{\mu\nu} e^{ikx}+e_{\mu\nu}^* e^{-ikx}
\end{equation}
with a wave vector $k^{\mu}$ ($\mu$=0,1,2,3) obeying the dispersion
relation derived from equation (\ref{eqpert})
\begin{equation}\label{kmukmu}
k_{\mu}k^{\mu}=0.
\end{equation}
The wave polarization $e_{\mu\nu}$ is a symmetric tensor whose
elements are tied with gauge conditions
\begin{equation}\label{cond3}
k_{\mu}e_{\nu}^{\mu}=0
\end{equation}
and
\begin{equation}\label{cond4}
e_{\mu}^{\mu}=0.
\end{equation}
We see that the familiar gravitational waves are also found in the
fourth order theory by using the covariance and the conformal
properties. The graviton  propagator behaves as $1/k^4$.

\subsection{Energy-momentum of gravitational waves}

The gravitational waves found in the last section were determined at
first order. However, the terms at higher order in the tensor
$W_{\mu\nu}$ can play the role of the energy-momentum tensor of
gravitational waves ($T_{\mu\nu}$ is still zero). We have
\begin{equation}\label{qw1}
W_{\mu\nu}=W_{\mu\nu}^{(0)}+W_{\mu\nu}^{(1)}+W_{\mu\nu}^{(2)}+...=0,
\end{equation}
with $W_{\mu\nu}^{(0)}=0$ (flat metric). Then
\begin{equation}\label{qw2}
W_{\mu\nu}^{(1)}=-W_{\mu\nu}^{(2)}-W_{\mu\nu}^{(3)}-...=-(W_{\mu\nu}-W_{\mu\nu}^{(1)})=\frac{1}{4\alpha}t_{\mu\nu},
\end{equation}
where $t_{\mu\nu}=-4\alpha (W_{\mu\nu}-W_{\mu\nu}^{(1)})$ is the
energy-momentum tensor carrying by the gravitational waves described
by $h_{\mu\nu}$. The first term in $t_{\mu\nu}$ is second order
in $h_{\mu\nu}$, so
\begin{eqnarray}\label{qw3}
t_{\mu\nu}^{(2)}=-4\alpha \Big
[-\frac{1}{6}\eta_{\mu\nu}\partial_{\sigma}\partial^{\sigma}
(\eta^{\lambda\rho}R_{\lambda\rho}^{(2)})+\frac{2}{3}\partial_{\mu}\partial_{\nu}(\eta^{\lambda\rho}R_{\lambda\rho}^{(2)})\nonumber\\
+\partial_{\sigma}\partial^{\sigma}
R_{\mu\nu}^{(2)}-\eta^{\beta\alpha}\partial_{\mu}\partial_{\alpha}R_{\beta\nu}^{(2)}-\eta^{\beta\alpha}\partial_{\nu}\partial_{\alpha}R_{\beta\mu}^{(2)}\Big
 ].
\end{eqnarray}
At second order in perturbation, the Ricci tensor takes the form
\begin{eqnarray}\label{r}
R_{\mu\nu}^{(2)}=\frac{1}{2}h^{\lambda\alpha}(\partial_{\mu}\partial_{\nu}h_{\lambda\alpha}
-\partial_{\nu}\partial_{\lambda}h_{\mu\alpha}
-\partial_{\mu}\partial_{\alpha}h_{\lambda\nu}
+\partial_{\alpha}\partial_{\lambda}h_{\mu\nu})\nonumber\\
+\frac{1}{4}(\partial_{\lambda}h_{\sigma\nu}+\partial_{\nu}h_{\sigma\lambda}-\partial_{\sigma}h_{\lambda\nu})
(\partial^{\lambda}h_{\mu}^{\sigma}
+\partial_{\mu}h^{\sigma\lambda}-\partial^{\sigma}h_{\mu}^{\lambda})\nonumber\\
-\frac{1}{4}(2\partial_{\alpha}h_{\sigma}^{\alpha}
-\partial_{\sigma}h)(\partial_{\nu}h_{\mu}^{\sigma}
+\partial_{\mu}h_{\nu}^{\sigma} -\partial^{\sigma}h_{\mu\nu}).
\end{eqnarray}
Using gauge conditions (\ref{cond1}) and (\ref{cond2}),
$R_{\mu\nu}^{(2)}$ becomes
\begin{equation}\label{rordre2}
R_{\mu\nu}^{(2)}=-\frac{3}{4}k_{\mu}k_{\nu}e^{\lambda\rho}e_{\lambda\rho}e^{2ikx}
-\frac{3}{4}k_{\mu}k_{\nu}e^{\lambda\rho}e_{\lambda\rho}^{\ast}+h.c.
\end{equation}
Inserting it in (\ref{qw1}) we find that
\begin{equation}\label{zero}
t_{\mu\nu}^{(2)}=0,
\end{equation}
the energy-momentum tensor of gravitational waves vanishes at the
lowest order.

\section{Domain wall solitons}
It has been observed that the theory of conformal gravity with a conformally coupled scalar field can be transformed by an appropriate field redefinition to an Einstein theory of gravity with coupled matter fields and a decoupled conformal sector which does not interact with the observable fields \cite{f}.  This analysis critically depends on the non-vanishing of the scalar field.  Such a situation might be locally valid, however, globally it is certainly not necessary.  We are reminded of an analogous situation concerning gauge fixing.  Consider the Georgi-Glashow model, which is a theory with a triplet scalar field and non-abelian local gauge symmetry $SO(3)$ that is spontaneously broken to $U(1)$ \cite{gg}.  In unitary gauge, one takes the the scalar field to point in the third direction, and then looking at the second order Lagrangian we can read off the spectrum, a theory of a massless $U(1)$ gauge boson, two massive vector gauge bosons and one massive neutral scalar, indeed the model was first invented in the early 70's as an alternative to the Weinberg-Salam model since the evidence for the massive Z vector boson was not conclusive at that time.  This is a perfectly fine analysis as long as the scalar field does not vanish.  However there exist configurations which require zeros of the scalar field.  Then the presumed choice of gauge becomes meaningless.  These are the 'tHooft-Polyakov magnetic monopole configurations.  The existence of these configuration can radically affect the physical spectrum of the theory.  In 2+1 dimensions, for example, the magnetic monopole play the role of instantons, and in fact cause the abelian $U(1)$ gauge theory to be confining \cite{polyakov}.  Insisting that the scalar field never vanish can throw out much more than expected.  Indeed, soliton type solutions in conformal gravity have already been found, where it is required that the scalar field vanish \cite{efp}.

In this same spirit, we look for domain wall configurations that require the Higgs field to vanish.    In both symmetry breaking vacuum type solutions that we have found that the scalar field satisfies
$$
\varphi_0^2=\frac{R}{24\lambda}
$$
or equivalently
$$\varphi_0=\pm\sqrt{\frac{R}{24\lambda}}.
$$
Thus a configuration which is $\varphi_0=-\sqrt{\frac{R}{24\lambda}}$ in one region of space and $\varphi_0=+\sqrt{\frac{R}{24\lambda}}$ in another region of space must be separated by a domain wall, which also necessarily requires that the scalar field vanish at some point between the two regions. The location of the domain wall is nominally defined as the position of the zero of the scalar field.  The stability of the domain wall concerns two different analyses.  First the domain wall may be of finite or infinite transversal extent.  An infinite domain wall can shrink in thickness to be a singular non-observable defect or expand in thickness to completely dilute itself through out space.  A finite domain wall must close on itself, forming a closed two surface embedded in three dimensional space.  Such a domain wall can have three possible instabilities.  It may shrink or expand in thickness as the infinite domain, but even if it is stable against these changes, it may also collapse and shrink to a point.  We study closed domain wall configurations in the maximally symmetric vacuum solutions of subsection (\ref{mss}) and infinite domain walls configurations in the solutions of subsection (\ref{nmss})
\subsection{Spherical domains in maximally symmetric space-times}
\begin{equation}
d\tau^2=(1+B(r))dt^2-\frac{1}{(1+B(r))}dr^2-r^2d\theta^2-r^2\sin^2(\theta)d\phi^2\label{metric}
\end{equation}
which includes anti-de Sitter space-time if $B(r)=k r^2$.  We then try a spherical ansatz for the scalar field, that the scalar field is in the vacuum configuration $\varphi=-\sqrt{\frac{R}{24\lambda}}$ for $r\ll r_0$ and $\varphi=+\sqrt{\frac{R}{24\lambda}}$ for $r\gg r_0$.  The scalar field will interpolate between the two vacua at some point near $r_0$.  It is evident that such a configuration, even if it is stable under scaling of its local thickness, will minimize its action by shrinking towards a point.  What happens when the nontrivial scalar field configurations from opposite sides of the spherical domain wall start to feel each others presence is not a priori obvious.  Here we show, for $\lambda$ large enough,  that the spherical bubble is unstable to shrinking to a point and to disappear.  We do this by considering a scale transformation on the field $B(r)$ and the field $\varphi(r)$ and observe its effects on the action.  It is important here to use the action as given by equations (\ref{weylaction1}) or (\ref{weylaction2}), as we are dealing with asymptotically maximally symmetric space-times for which these two expressions vanish but not the one given in equations (\ref{weylaction3}).  With the metric of equation (\ref{metric}) it is a straightforward, although a little tedious, calculation to find the non-zero components of the Riemann tensor.  We find the non-zero components, written in tangent space indices
$$
R^0_{~101}=-\frac{B''}{2},\quad R^0_{~202}=R^0_{~303}=-\frac{B'}{2r}=R^1_{~212}=R^1_{~313}\quad R^2_{~323}=-\frac{B}{r^2}
$$
where the corresponding orthonormal basis of one-forms of the (dual) tangent space  are given by $\sigma^0=-\sqrt{1+B}dt$, $\sigma^1=\frac{1}{\sqrt{1+B}}dr$, $\sigma^2=rd\theta$, and $\sigma^3=r\sin(\theta)d\phi$.   Then the gravitational action is given by 
\begin{eqnarray}
I_W&=&-\alpha\int d^4x~ r^2\sin \theta\left( \left(B''^2+\frac{4B'^2}{r^2}+\frac{4B^2}{r^4}\right)\right.\nonumber\\ & &-2\left(2\left(\frac{B''}{2}+\frac{B'}{r}\right)^2+2\frac{(rB)'^2}{r^4}\right) +\frac{1}{3}~4\left.\left(\frac{(r^2B)''}{r^2}\right)^2\right)\label{ga}
\end{eqnarray}
while the scalar field action is given by
\begin{equation}\label{scalarfield}
I_{M}=\int d^{4}x~r^2\sin \theta \Big(- \frac{1}{2}(1+B(r))\partial_{r}\varphi\partial_{r}\varphi +\frac{1}{12}\frac{(r^2B)''}{r^2}\varphi^2 -\lambda\varphi^4 \Big),\label{ma}
\end{equation}
where we have already specialized to static, spherically symmetric configurations.  It is easy to show for sufficiently large $\lambda$ or conversely, sufficiently large $\alpha$, this action is made up of a sum of individually negative definite terms.  Action (\ref{ga}) is by itself re-expressible as such a sum, and completing the square in the action (\ref{ma}) yields an additional  term of the form $(1/\lambda )((r^2B)''/{ r^2})^2$ which is dominated by the corresponding term in the action (\ref{ga}) for sufficiently small $\lambda$.   In fact a constant needs to be added to the action,  chosen to ensure that the action vanishes for the vacuum configuration.  Indeed the action is otherwise an infinite constant for the vacuum configuration.

Using a modified scaling argument that this action does not have a stable stationary point except for the trivial vacuum.  Consider the replacement:
\begin{eqnarray}
\varphi(x)\rightarrow\varphi(\Lambda x)\\
B(x)\rightarrow\frac{B(\Lambda x)}{\Lambda^2}
\end{eqnarray}
Under this transformation, $\varphi (x)$ is squeezed into an even smaller region of space as $\Lambda\rightarrow\infty$, although the constant value of $\varphi$ as is attained asymptotically, does not change in magnitude.  Correspondingly, $B(x)$ is also squeezed and reduced in magnitude, however $B''(x)\rightarrow \left.\frac{\partial^2B(y)}{\partial y^2}\right|_{y=\Lambda x}$ does not change in magnitude.  Since the curvature scalar is controlled by $B''(x)$ and dimensionally equivalent objects, the value of the scalar fields symmetry breaking minimum is not altered by such a rescaling.  Then under this rescaling we find
\begin{eqnarray}
I_W&=&-\alpha\int d^4x r^2\sin \theta {\cal L}(r)\rightarrow -\alpha\int d^4x r^2\sin \theta {\cal L}(\Lambda r)\nonumber\\
&=&-\frac{\alpha}{\Lambda^3}\int dt~d\theta ~d\phi ~d(\Lambda r) (\Lambda r)^2\sin \theta ~{\cal L}(\Lambda r)=I_W/\Lambda^3
\end{eqnarray}
and
\begin{eqnarray}
I_{M}\rightarrow\int d^{4}x~r^2\sin \theta \Big(- \frac{1}{2}(1+B(\Lambda r)/\Lambda^2)\partial_{r}\varphi\partial_{r}\varphi(\Lambda r)\nonumber\\
+\frac{1}{12}\frac{(r^2B(\Lambda r)/\Lambda^2)''}{r^2}\varphi^2(\Lambda r)-\lambda
\varphi^4(\Lambda r) \Big)\nonumber\\
=\int d^{4}x~r^2\sin \theta  \Big(- \frac{1}{2}(1+B(\Lambda r)/\Lambda^2)\Lambda^2\left(\left.\partial_{y}\varphi(y)\partial_{y}\varphi(y)\right|_{y=\Lambda r}\right)\nonumber\\
+\frac{1}{12}\frac{1}{\Lambda^2r^2}\left.\frac{\partial^2(y^2B(y)}{\partial y^2}\right|_{y=\Lambda r}\varphi^2(\Lambda r)-\lambda
\varphi^4(\Lambda r) + const.\Big)\nonumber\\
=\frac{1}{\Lambda^3}\int d^{4}x~r^2\sin \theta  \Big(- \frac{1}{2}(1+B( r)/\Lambda^2)\Lambda^2\left(\partial_{r}\varphi(r)\partial_{r}\varphi(r)\right)\nonumber\\
+\frac{1}{12}\frac{1}{r^2}\frac{\partial^2(r^2B(r))}{\partial r^2}\varphi^2( r)-\lambda
\varphi^4( r) \Big) .
\end{eqnarray}
Thus we see that as $\Lambda\rightarrow\infty$, the whole classical action vanishes, belying the possibility of a non trivial stationary point of finite action.    We stress again that it is important that we used the expression (\ref{weylaction1}) or (\ref{weylaction2}) in this anaylsis, since (\ref{weylaction3}) is actually not finite for the configurations that we are considering.  Although this would mean that the spherical domain wall will necessarily collapse and shrink to a point, it does not mean that the bubble will be short lived.  Indeed the dynamics of the collapse could take a macroscopic amount of time, and the configurations could be most relevant to the physics of the theory.  

\subsection{Domain walls in non-maximally symmetric space-times}

In the solutions found in section (\ref{nmss}) we can insert an infinite planar domain wall and ask if the wall is stable against its width shrinking to zero or expanding to infinity.  Such a domain wall would, for a finite action configuration, have to close on itself say into a spherical bubble, and would probably collapse and shrink to a point.  But if the bubble is of a much larger radius that the thickness of its wall, then this collapse time can be quite long, and the domain wall configurations can be important for the dynamics of the theory.  

We take and ansatz of the form
\begin{equation}
d\tau^2=(1+B(x))dt^2-\frac{1}{(1+B(x)}dx^2-dy^2-dz^2
\end{equation}
where $B(x)$ is now a function that will interpolate between the two vacua, $\varphi_0=\pm\sqrt{\frac{B''(\pm\infty)}{24\lambda}}$, where $B''(\pm\infty)$ is a constant, and the critical coupling condition $\alpha=1/(192\lambda)$ is assumed to be satisfied.  Then the Riemann curvature is easily calculated with just two non-zero components
\begin{equation}
R^1_{~010}=B''(x)/2=-R^0_{~101}
\end{equation}
and the Ricci tensor also has just two non-zero components
\begin{equation}
R_{00}=B''(x)/2=-R_{11}
\end{equation}
finally giving the scalar curvature
\begin{equation}
R=B''(x).
\end{equation}
Since the space is flat in the $(y,z)$ directions, all curvature components with indices in these directions simply vanish.  The previous analysis for the equations of motion yields that the 11 a priori equations reduce to just two when the Bianchi identities (covariant conservation) and the tracefree nature of the tensors $W_{\mu\nu}$ and $T_{\mu\nu}$ are taken into account.  It is straightforward, but tediious to show this expllicitly, we will reproduce the calculation for the trace of $W_{\mu\nu}$.  The expression for $W_{\mu\nu}$ is given by:
\begin{equation}
W_{\alpha\beta}=\frac{1}{3}g_{\alpha\beta}\left(((1+B)B''')'-\frac{(B'')^2}{4}\right)+\frac{1}{3}(B'')_{;\alpha;\beta}
\end{equation}
with $(\alpha ,\beta )$ taking on the values $(0,1)$ while 
\begin{equation}
W_{ab}=\frac{-1}{6}g_{ab}\left(((1+B)B''')'-\frac{(B'')^2}{2}\right).
\end{equation}
where $(a,b)$ take on the values $(2,3)$.
Taking the trace yields
\begin{eqnarray}
W^\mu_{~~\mu}=\frac{2}{3}\left(((1+B)B''')'-\frac{(B'')^2}{4}\right)+\frac{1}{3}(B'')^{;\beta}_{~~;\beta}\nonumber\\
+\frac{-1}{3}\left(((1+B)B''')'-\frac{(B'')^2}{2}\right).
\end{eqnarray}
Since
\begin{equation}
(B'')^{;\beta}_{~~;\beta}=\frac{1}{\sqrt{-g}}\partial_\beta \sqrt{-g}g^{\beta\alpha}\partial_\alpha B''= \partial_\beta g^{\beta\alpha}\partial_\alpha B''=-((1+B)B''')',
\end{equation}
it is evident that the trace vanishes.   

Thus the equations left to satisfy are
\begin{eqnarray}
-2\alpha W_{11}=T_{11}/2\hskip3cm\\
(1+B(x))\varphi''(x)+B'(x)\varphi'(x)+\frac{1}{6}B''(x)\varphi(x)-4\lambda\varphi^3(x)=0.
\end{eqnarray}
These reduce to
\begin{eqnarray}
\frac{2\alpha}{3}\left(B'''B'-\frac{1}{2}(B'')^2\right)=\frac{1}{2}(1+B)(\varphi')^2+\frac{1}{6}\varphi B'\varphi'-\lambda\varphi^4\\
(1+B(x))\varphi''(x)+B'(x)\varphi'(x)+\frac{1}{6}B''(x)\varphi(x)-4\lambda\varphi^3(x)=0.
\end{eqnarray}
We can verify, that for critical coupling, $\alpha=1/(192\lambda )$, we get the vacuum solution $B=(R/2)r^2$, $R$ a constant,  $\varphi_0=\pm\sqrt{R/24\lambda}$.  

The further analysis of these equations requires numerical methods.  We have managed to find profiles of approximate solutions of the equations using numerical relaxation methods.  The configuration is initially taken to be 
\begin{eqnarray}
\varphi(x)=\sqrt{\frac{R}{24\lambda}}\tanh(x)\\
B(x)=1+\frac{R}{2}x^2
\end{eqnarray}
which asymptotically, for $x\rightarrow\pm\infty$, attains the vacuum-like field configuration. 
The initial configuration is allowed to relax with a dissipative linear term.  The equations considered are given by
\begin{eqnarray}
k\dot B+\frac{2\alpha}{3}\left(B'''B'-\frac{1}{2}(B'')^2\right)=\frac{1}{2}(1+B)(\varphi')^2+\frac{1}{6}\varphi B'\varphi'-\lambda\varphi^4\\
k\dot\varphi+(1+B(x))\varphi''(x)+B'(x)\varphi'(x)+\frac{1}{6}B''(x)\varphi(x)-4\lambda\varphi^3(x)=0.
\end{eqnarray}
where now $B=B(x,t)$ and $\varphi=\varphi(x,t)$.  With appropriate choices for the coefficients, we obtain configurations which seem to have relaxed to a domain wall.  Below in  Figure (\ref{figf}) is a graph of $\varphi(x)$ and in Figure (\ref{figb}) is a graph of $B(x)$ that we have obtained via numerical relaxation.  

\begin{figure}[ht]
  \hfill
  \begin{minipage}[t]{.45\textwidth}
    \begin{center}  
      \includegraphics[scale=.5]{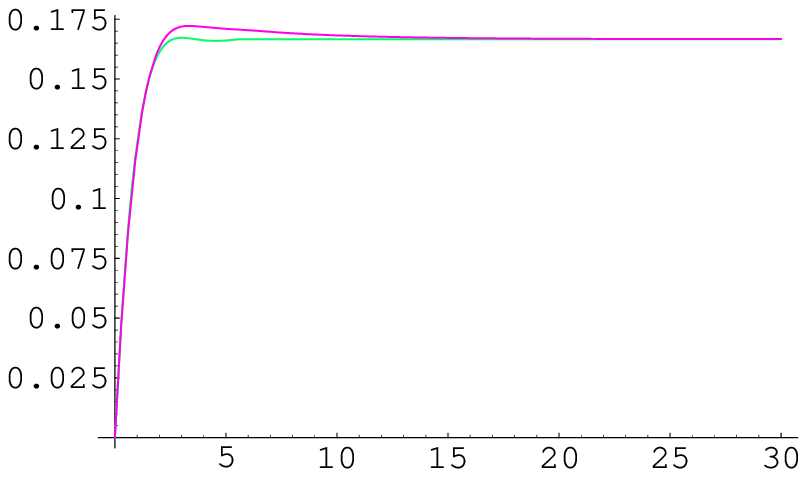}
      \caption{$\varphi(x)$ for $x\in [\sim 0,30]$, initial (green) and final (purple). }
      \label{figf}
    \end{center}
  \end{minipage}
  \hfill
  \begin{minipage}[t]{.45\textwidth}
    \begin{center}  
      \includegraphics[scale=0.5]{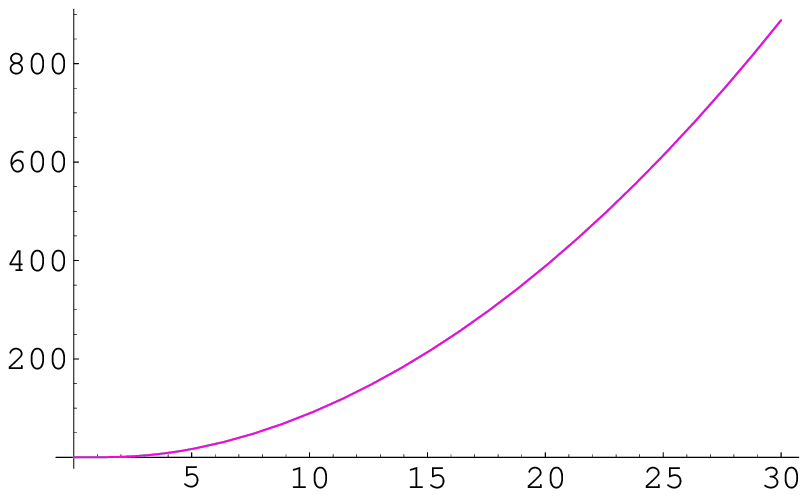}
      \caption{$B(x)$ for $x\in [\sim 0,30]$, initial (green) and final (purple) (they are coincident within the resolution).}
      \label{figb}
    \end{center}
  \end{minipage}
  \hfill
\end{figure}

\section{Conclusions}
We have studied the  spontaneously breaking of the conformal symmetry in Weyl gravity, conformally coupled to scalar matter.  The spontaneous symmetry breaking gives mass to the matter field and chooses a solution that is asymptotically anti-de Sitter for the geometry.  We have shown that gravitational waves, to second order in perturbation do not carry any energy or momentum.  We have found numerical evidence that there exist configurations, which are most likely long lived, where the topology requires that the scalar field have a zero. This implies that there can be no conformal transformation that can remove the scalar field, and relegate the conformal invariance to an unobservable disjoint sector, contrary to what  has been recently suggested \cite{f}.  We feel that there is much motivation to study this theory in greater detail as it may truly be an alternative theory to Einsteinian gravitation, that does not carry the burden of having to explain missing gravitational forces with huge amounts of dark matter and energy. 

\acknowledgments
We thank O.P.S. Negi for useful discussions.  We also thank the (Kavli) Institute for Theoretical Physics of the Chinese Academy of Sciences, Beijing, where this work was completed, for financial support and pleasant working conditions.  This work is supported  by NSERC of Canada, the Tunisian Ministry of Education with a graduate fellowship and  the Center for Quantum Spacetime of Sogang University with grant number R11-2005-021   for financial support.

\end{document}